%% file: main.tex
\documentclass[sigconf, review=false, noacm]{acmart}

\AtBeginDocument{%
  }
\usepackage{subcaption}
\usepackage{graphicx}
\usepackage{cleveref}

\crefname{figure}{Figure}{Figures}
\Crefname{figure}{Figure}{Figures}
\crefname{subfigure}{Figure}{Figures}
\Crefname{subfigure}{Figure}{Figures}

\copyrightyear{2024}
\acmYear{2024}
\setcopyright{rightsretained}
\acmConference[CIKM '24]{Proceedings of the 33rd ACM International Conference on Information and Knowledge Management}{October 21--25, 2024}{Boise, ID, USA}
\acmBooktitle{Proceedings of the 33rd ACM International Conference on Information and Knowledge Management (CIKM '24), October 21--25, 2024, Boise, ID, USA}
\acmDOI{10.1145/3627673.3679223}
\acmISBN{979-8-4007-0436-9/24/10}

\begin{document}

\title{Music2P: A Multi-Modal AI-Driven Tool for Simplifying Album Cover Design}


\author{Joong Ho Choi}
\affiliation{%
  \institution{Carnegie Mellon University}
  \city{Pittsburgh}
  \country{USA}}
\email{joonghoc@andrew.cmu.edu}

\author{Geonyeong Choi}
\affiliation{%
  \institution{Handong Global University}
  \city{Pohang}
  \country{South Korea}}
\email{gychoi@handong.ac.kr}

\author{Ji-Eun Han}
\affiliation{%
  \institution{KT}
  \city{Seoul}
  \country{South Korea}}
\email{ji-eun.han@kt.com}

\author{Wonjin Yang}
\affiliation{%
  \institution{UNIST}
  \city{Ulsan}
  \country{South Korea}}
\email{wonjinyang@unist.ac.kr}
\author{Zhi-Qi Cheng}
\affiliation{%
  \institution{Carnegie Mellon University}
  \city{Pittsburgh}
  \country{USA}}
\email{zhiqic@cs.cmu.edu}
\renewcommand{\shortauthors}{Choi et al.}


\input{section/0_abstract}   

\keywords{Multi-modal, Album cover design, AI-assisted creativity}
\maketitle

\input{section/1_introduction}

\input{section/2_method}
\input{section/3_demo}
\input{section/4_conclusion}

\bibliographystyle{ACM-Reference-Format}
\bibliography{sample-base}

\end{document}

%% file: section/0_abstract.tex
\begin{abstract}
In today's music industry, album cover design is as crucial as the music itself, reflecting the artist's vision and brand. However, many AI-driven album cover services require subscriptions or technical expertise, limiting accessibility. To address these challenges, we developed Music2P, an open-source, multi-modal AI-driven tool that streamlines album cover creation, making it efficient, accessible, and cost-effective through Ngrok. Music2P automates the design process using techniques such as Bootstrapping Language Image Pre-training (BLIP), music-to-text conversion (LP-music-caps), image segmentation (LoRA), and album
cover \& QR code generation (ControlNet). This paper demonstrates the Music2P interface, details our application of these technologies, and outlines future improvements. Our ultimate goal is to provide a tool that empowers musicians and producers, especially those with limited resources or expertise, to create compelling album covers.
\end{abstract}

%% file: section/1_introduction.tex
{\def\thefootnote{}\footnotetext{This project by Joong Ho Choi, Geonyeong Choi, Ji-Eun Han, and Wonjin Yang was completed in Fall 2023 for the 11775: Large Scale Multimedia Analytics course at Carnegie Mellon University, instructed by Dr. Zhi-Qi Cheng.}}

\section{Introduction}
In the contemporary music industry, album cover design plays a pivotal role, serving as a visual embodiment of an artist's vision and brand. Traditionally, this process has required collaborative efforts among musicians, graphic designers, and marketing teams \cite{dorochowicz2019relationship}. However, independent artists and smaller labels face significant challenges due to limited time, resources, and specialized design expertise. The key challenge is creating a visual representation that captures the music's essence, resonates with the target audience, and stands out in a competitive market—all within tight time and budget constraints.

\begin{figure}[!t]
\centering
\vspace{1.5em}
\includegraphics[width=0.9\linewidth]{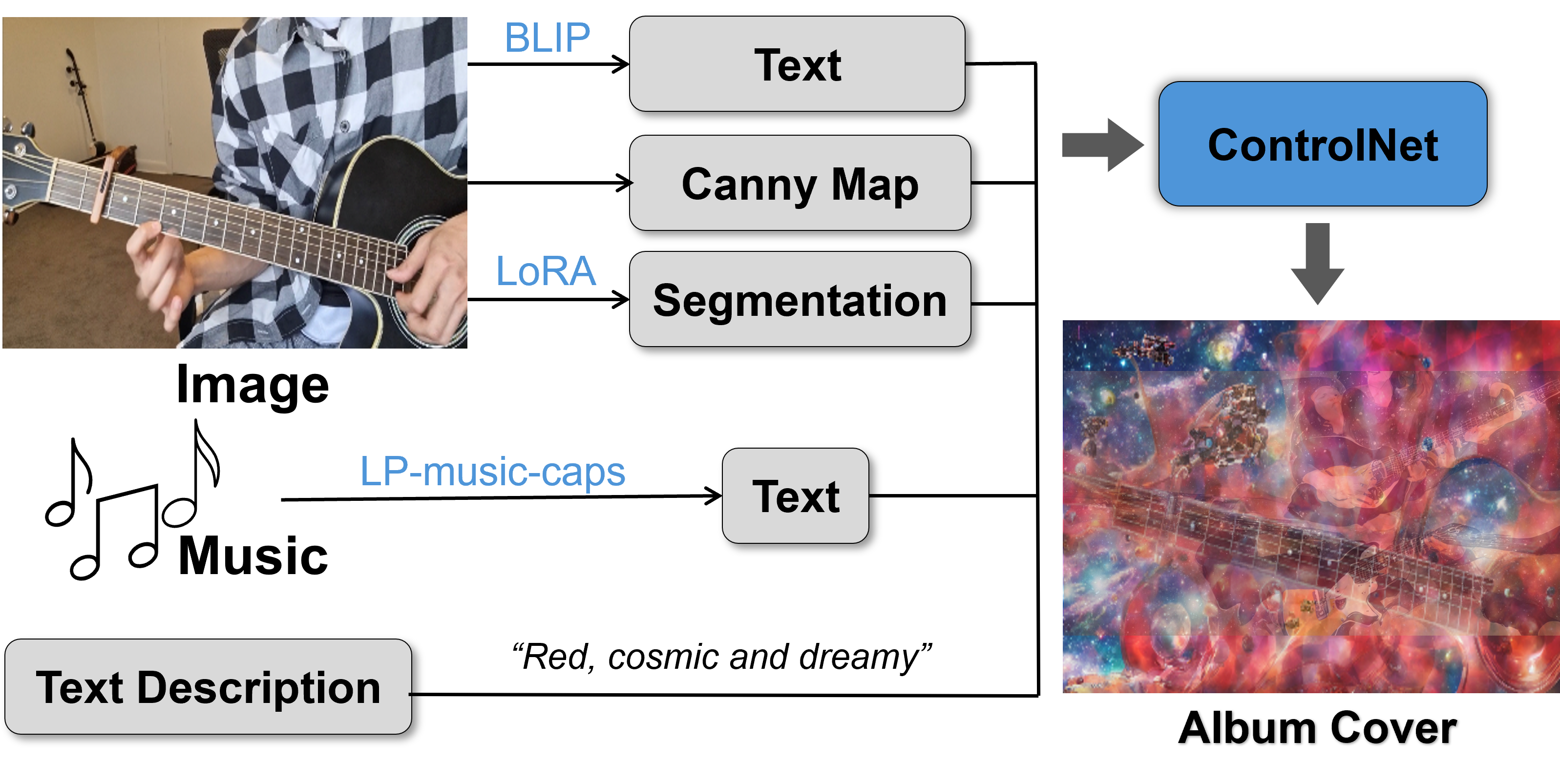}
\vspace{-1em}
\caption{Overview of the Music2P tool for album cover generation. The system integrates Bootstrapping Language Image Pre-training (BLIP), music-to-text conversion (LP-music-caps), image segmentation (LoRA), and ControlNet with QR code generation. These components process multi-modal inputs—text, image, and audio—producing visually and contextually appropriate album covers.}
\vspace{-1em}
\label{fig:arch}
\end{figure}

In response to these challenges, recent years have witnessed the emergence of AI-driven solutions for album cover generation~\cite{hepburn2017album,marien2022audio,campobadal2023ai}.~However, these solutions often suffer from uni-modal constraints, typically accepting only textual input. This limitation restricts users' creativity and the breadth of information they can provide \cite{hanna2023use}. Furthermore, due to high computational costs, these services often limit users to a small number of attempts, leaving independent artists and smaller labels with minimal opportunities to generate satisfactory album covers.

To address these limitations, we introduce Music2P\footnote{Project available at \url{https://github.com/JC-78/Music2P}}, a system that leverages recent AI technologies to provide a more comprehensive and adaptable solution. As illustrated in Figure \ref{fig:arch}, Music2P integrates multiple components to create an attractive album cover generation pipeline. The Music2P's system architecture includes image captioning (BLIP \cite{LiEtAl2022BLIP}), music-to-text conversion (LP-music-caps \cite{lewis-etal-2020-bart}), image segmentation (LoRA \cite{hu2021lora}), album covers \& QR code generation (ControlNet \cite{ZhangAgrawala2023}). These interconnected components work together to produce aesthetically pleasing and semantically relevant album covers from multi-modal inputs, offering a holistic approach to the design process.

The key to Music2P's functionality lies in its multi-modal processing capabilities. Within its framework, we employ image processing techniques to extract textual content, edge features through Canny edge detection, and image segmentation to delineate thematic elements. Concurrently, we apply music analysis algorithms to distill lyrical content from instrumental and vocal compositions. This multi-modal information, combined with user-provided textual descriptors, is fed into ControlNet \cite{ZhangAgrawala2023}, a neural network architecture designed to control image generation through task-specific conditions. As shown in Figure \ref{fig:arch}, ControlNet integrates and interprets these diverse data points, leveraging deep learning architectures to synthesize the extracted features into a cohesive visual output that reflects the music's intrinsic qualities. This process represents a convergence of technology and aesthetics, encapsulating the artist's narrative in a visual representation that enhances the listener's experience. By bridging the gap between audio, visual, and textual modalities, Music2P contributes to the ongoing discourse on the role of AI in artistic domains, offering an innovative alternative to traditional methodologies in visual music representation.

While maintaining high-quality album cover generation through the latest AI techniques, Music2P also addresses previously mentioned challenges by offering rapid and cost-effective service that anyone can easily deploy in minutes. Utilizing Ngrok \cite{ngrok}, a tool that facilitates secure connections between local development environments and the public internet, our solution can be deployed by users quickly. Additionally, Music2P offers a user-friendly interface where artists can upload their music, an image to use as the foundation or reference for album cover art, and a description of the desired tone or style to create visually compelling covers. This interactive approach is particularly beneficial for independent musicians and small music companies, leveling the playing field by providing high-quality design solutions that were previously inaccessible due to cost and technical barriers. Music2P thus stands as a revolutionary tool in the modern music industry, streamlining the album design process and empowering any artist to focus on their creative work.

To further enhance functionality, we have also included a QR-code generation feature, recognizing its common and effective use in promotional strategies. In 2022, two noteworthy incidents highlighted the effectiveness of QR codes. During Super Bowl LVI, CoinBase displayed a QR code on the screen, attracting numerous scans and succeeding as a marketing campaign \cite{coinbase_ad}. Additionally, in June 2022, aesthetic AI-generated QR codes began appearing on Reddit. Inspired by these events and considering the financial limitations of our intended users, we incorporated an AI QR-code generation service into our tool.

To our knowledge, there are currently limited open-source projects that offer a multi-modal solution for album cover generation specifically oriented towards independent musicians and small labels. Overall, the contributions of Music2P are summarized as follows:
\begin{itemize}
\item {Music2P provides automatic cost-effective,end-to-end album cover generation with user-given multi-modal inputs.}
\vspace{0.5em}
\item {Music2P can generate a shareable QR code with the album cover generated by our system.}
\end{itemize}

%% file: section/2_method.tex
\begin{figure}[t]
  \centering
  \begin{minipage}[t]{0.27\linewidth}
    \centering
    \includegraphics[width=\linewidth]{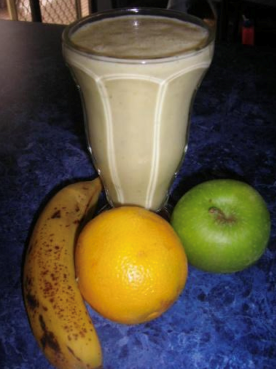}
    \subcaption{\small Original Image}
    \label{fig:original_canny}
  \end{minipage}
  \hspace{0.02\linewidth}
  \begin{minipage}[t]{0.27\linewidth}
    \centering
    \includegraphics[width=\linewidth]{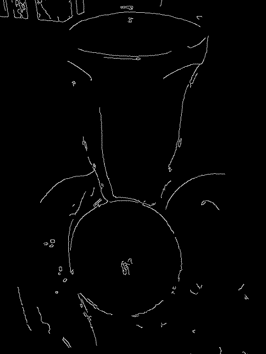}
    \subcaption{\small Canny Edge Detection}
    \label{fig:canny_det}
  \end{minipage}
\hspace{0.02\linewidth} 
  \begin{minipage}[t]{0.27\linewidth}
    \centering
    \includegraphics[width=\linewidth]{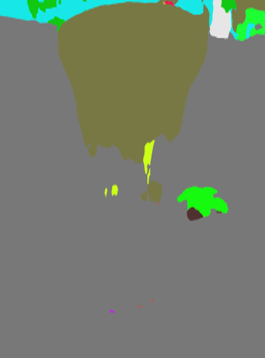}
    \subcaption{\small Image Segmentation}
    \label{fig:fruit_seg}
  \end{minipage}
  \vspace{0.02\linewidth}
  \begin{minipage}[t]{0.43\linewidth}
    \centering
    \includegraphics[width=\linewidth]{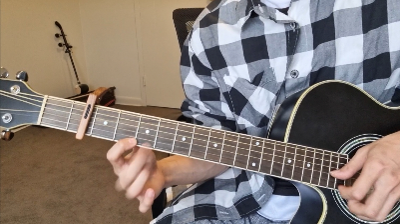}

    \subcaption{\small Original Image}
    \label{fig:original_seg}
  \end{minipage}
  \hspace{0.02\linewidth}
  \begin{minipage}[t]{0.43\linewidth}
    \centering
    \includegraphics[width=\linewidth]{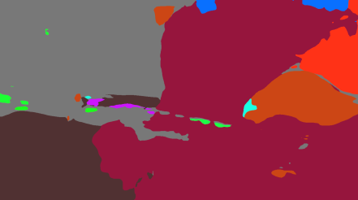}
    \subcaption{\small Image Segmentation}
    \label{fig:img_seg}
  \end{minipage}
  \vspace{-1.5em}
  \caption{Comparison of original image, Canny edge detection, and image segmentation.~[Zoom in to view]}
  \label{fig:comp}
  \vspace{-1.5em}
\end{figure}

\section{Music2P: \textit{Multi-Modal Album Cover Generation}}
As illustrated in Figure \ref{fig:arch}, Music2P is an open-source, multi-modal AI system designed to streamline the creation of personalized album covers by integrating various input modalities, including text, image, and audio. This system addresses accessibility challenges in the music industry by offering a cost-effective, subscription-free solution. Music2P automates the album cover design process using advanced techniques, including Bootstrapping Language Image Pre-training (BLIP) \cite{LiEtAl2022BLIP} for image captioning, LP-music-caps \cite{doh2023lpmusiccaps} for converting music to text, image segmentation with Low-Rank Adaptation (LoRA) \cite{hu2021lora}, and ControlNet \cite{ZhangAgrawala2023} for generating album covers \& QR codes.

\subsection{Image to Text Conversion}
We use Bootstrapping Language-Image Pre-training (BLIP) as a composite caption generator for images. It has an advanced encoder-decoder structure that can handle multi-modal tasks. Specifically, BLIP uses self-attention and cross-attention to improve the understanding of the given input image for album cover design. BLIP also includes a filter to remove noisy or unrelated captions, allowing us to use the most appropriate and influential descriptions as input when creating album covers.

\subsection{Music to Text Conversion}
We employ LP-music-caps \cite{doh2023lpmusiccaps} as its Audio-to-Caption model, a sophisticated system designed to transform musical input into descriptive text. This model's development follows a two-phase approach: initial training on a GPT 3.5 Turbo-generated pseudo dataset of music-caption pairs, followed by fine-tuning with human-crafted captions to enhance output naturalness.
LP-music-caps generates segmented captions at ten-second intervals, offering nuanced insights into the music's evolving themes, moods, and expressions. This sequential captioning method effectively captures the musical piece's dynamic and temporal progression, translating auditory nuances into textual form.

To synthesize these multiple captions into a cohesive narrative, Music2P utilizes BART, a large-scale model fine-tuned on the CNN/Daily Mail dataset, as its Summarizer. BART employs abstractive summarization techniques to distill key information from the LP-music-caps-generated captions. The result is a concise, comprehensive caption that encapsulates the entire musical piece's essence, serving as a crucial input for the album cover generation process.
This Music to Text conversion acts as a critical bridge across audio, textual, and visual modalities. To demonstrate its efficacy, we conducted an experiment using a composed piece\footnote{The composed song can be found at: \url{https://youtu.be/0MkPrE3Mfmw}}. 

Our model accurately captured the one-minute guitar-based song's essence with the following caption: "The low-quality recording features a live performance of a pop song that consists of a passionate male vocal singing over acoustic guitar arpeggiated chords. The recording is noisy and in mono."
This music-to-text conversion not only enhances the multi-modal capabilities of Music2P but also ensures that the generated album covers are deeply rooted in the musical content they represent. By effectively translating auditory elements into textual descriptions, we create a robust foundation for the visual interpretation of music in the form of album artwork.

\subsection{Segmentation with LoRA}
In album cover creation, extracting a Canny edge map and performing image segmentation on user-provided images are crucial for aligning visual design with musical aesthetics and themes. The Canny map provides a simplified edge-based abstraction, but its monochromatic nature inherently limits its ability to differentiate color-defined objects, as evidenced in \cref{fig:canny_det}.
This limitation becomes critical in contexts requiring color differentiation. The Canny map's reduction of images to binary edge representations results in the loss of vital chromatic information essential for object identification and differentiation. Conversely, image segmentation algorithms excel in parsing images into distinct segments based on color, intensity, and texture, as demonstrated in \cref{fig:img_seg}.

Both Canny detection and image segmentation, despite their individual strengths, exhibit notable limitations. Canny detection effectively identifies edges (\cref{fig:canny_det}) akin to the original image (\cref{fig:original_canny}) but fails in color distinction, rendering it ineffective for differentiating objects like apples and oranges. Image segmentation faces similar challenges when applied to images outside its training domain, as illustrated by the ineffective segmentation resulting from applying a cityscape-trained model to a fruit image (\cref{fig:fruit_seg}).

Addressing these limitations necessitates fine-tuning the segmentation network, a process involving the adaptation of a pre-existing network with approximately 90 million parameters. To manage this computational challenge efficiently, we employed Low-Rank Adaptation (LoRA) \cite{hu2021lora}. LoRA enables efficient retraining of large neural networks by adjusting only a small parameter subset, significantly reducing computational demands and training time.
The LoRA technique is mathematically expressed as:
\begin{equation}
h = W_0x + \Delta Wx = W_0x + BAx,
\end{equation}
where the hidden state $h$ comprises the original transformation $W_0x$ and a low-rank update $\Delta Wx$, decomposed into the product of smaller matrices $B \in \mathbb{R}^{d \times r}$ and $A \in \mathbb{R}^{r \times k}$, with rank $r$ substantially smaller than dimensions $d$ and $k$. This low-rank approximation facilitates efficient tuning of large models through updates to a small parameter subset.

\begin{figure}[t]
  \centering
  \includegraphics[width=0.9\linewidth]{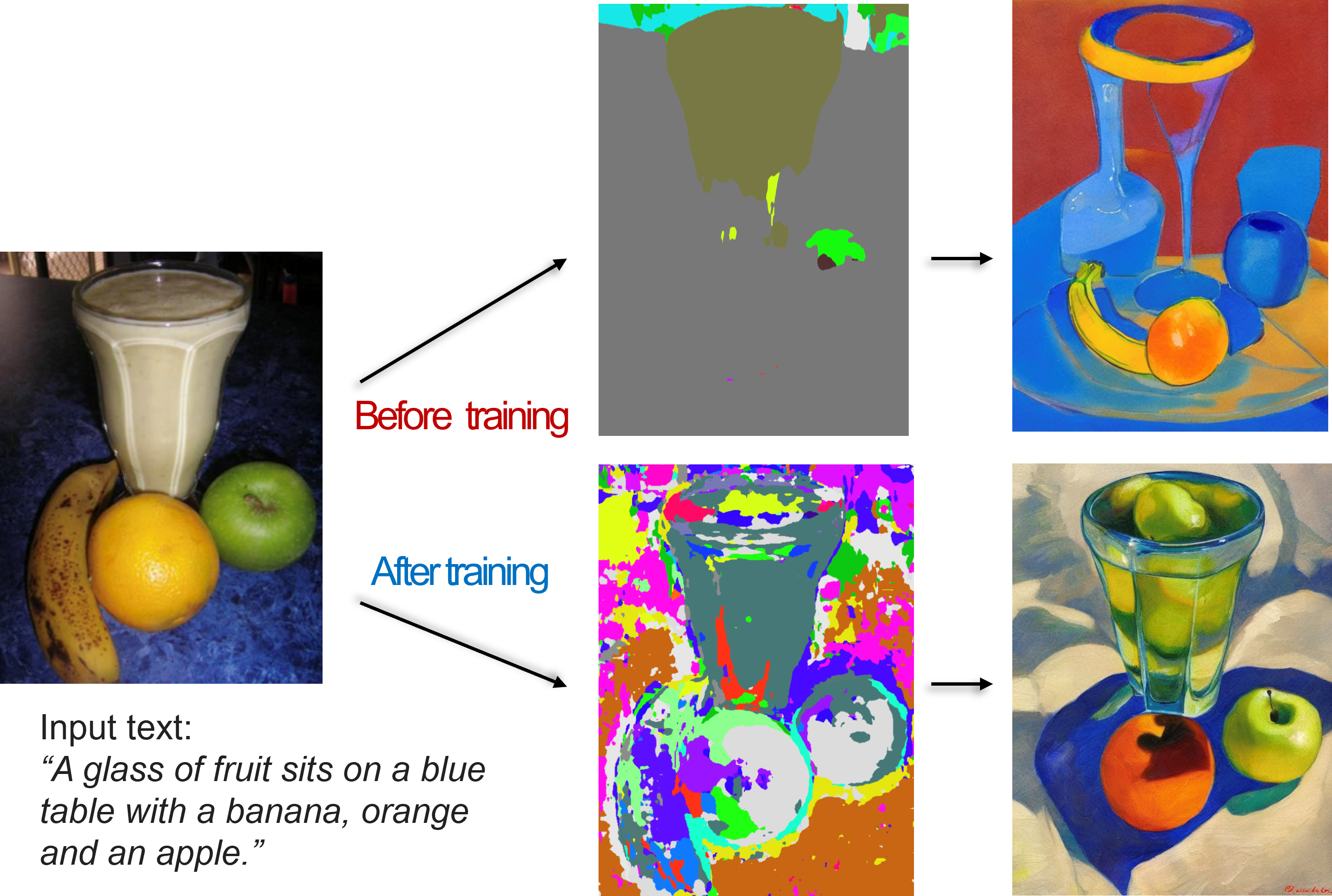}
  \vspace{-0.5em}
  \caption{Comparative analysis of images generated before and after LoRA model training.~[Zoom in for detail]}
  \vspace{-1em}
  \label{fig:seg_and_LoRA}
\end{figure}

Applying LoRA, we successfully trained Music2P's segmentation network on a food dataset, markedly improving its ability to accurately segment food items, as illustrated in \cref{fig:seg_and_LoRA}.
This enhanced segmentation capability enables the network to recognize diverse food-related elements, crucial for generating contextually relevant album covers for music with culinary themes. The fine-tuned segmentation network thus becomes an integral component of Music2P, facilitating the creation of visually rich and thematically coherent album artwork through effective interpretation and processing of diverse image content.

\subsection{Album Cover Generation using ControlNet}
ControlNet \cite{ZhangAgrawala2023}, a suite of neural networks fine-tuned on Stable Diffusion, forms the core of Music2P's image generation. Its architecture seamlessly integrates pretrained parameters from a Diffusion model—specifically, the latent space of Stable Diffusion's U-Net architecture—with separately maintained pre-trained parameters as locked copies. This unique design enables precise control over both structural and artistic elements in the generated images. Specifically, the ControlNet methodology employs a dual-copy strategy, creating two versions of a large image diffusion model: one trainable and one with frozen weights. This approach serves a dual purpose: preserving the foundational knowledge encoded in the Diffusion model's latent space while allowing adaptability through the trainable copy. "Zero convolution" layers bridge these two copies, facilitating seamless information flow and enabling network fine-tuning to meet the specific demands of album cover design.

This architecture empowers ControlNet to perform task-specific conditioning, significantly enhancing control over various aspects of image generation, including pose, edge detection, and depth maps. For Music2P, we further refined the pre-trained ControlNet through additional fine-tuning, focusing on datasets for edge detection and image segmentation, tailoring it specifically for album cover creation.
To generate album covers, Music2P integrates ControlNet by leveraging a triad of inputs:
\begin{enumerate}
    \item BLIP-generated captions derived from the input image
    \item Textual descriptions extracted from the audio file via Music to Text conversion
    \item User-provided prompts for additional creative direction
\end{enumerate}
This multi-modal approach enables ControlNet to process a rich tapestry of inputs—encompassing textual themes, visual elements, and musical mood. The result is album covers that transcend mere visual representation, becoming holistic echoes of the music's essence, the artist's intent, and the image's thematic content.

By synthesizing these diverse inputs, ControlNet ensures each generated album cover is not only visually striking but also deeply resonant with the music it represents. This comprehensive methodology yields album artwork that is both aesthetically pleasing and semantically aligned with the artist's vision, effectively bridging the gap between audio and visual mediums in a way that traditional design processes often struggle to achieve.

\subsection{QR Code Generation}
\label{subsec:qr_code_generation}
\begin{figure}[t]
  \centering
  \begin{minipage}[t][5cm][t]{0.4\linewidth} 
    \centering
    \includegraphics[width=3cm, height=3cm]{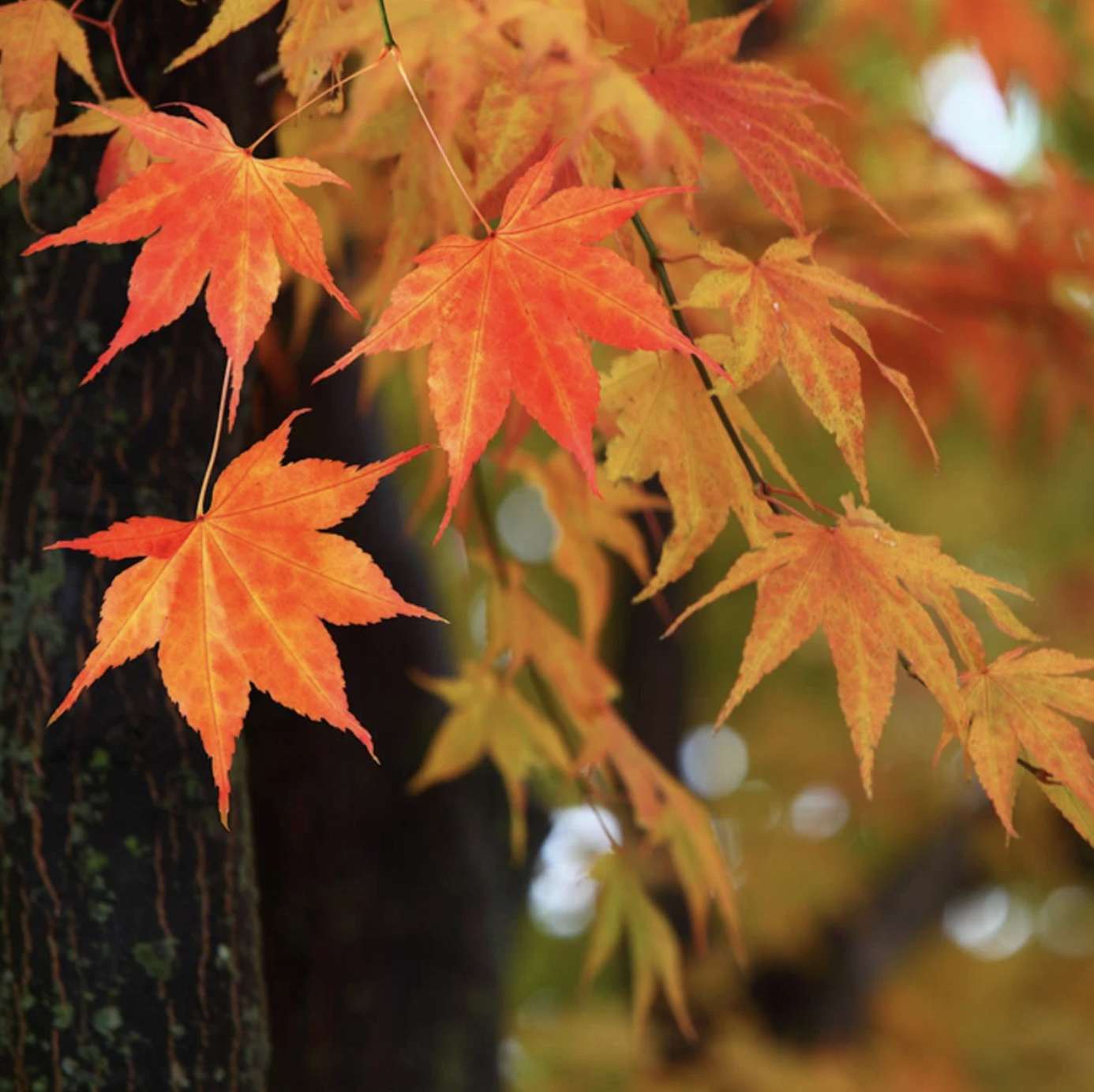}
    \subcaption{\small User-provided image}
    \label{fig:4a}
  \end{minipage}
  \hspace{0.02\linewidth} 
  \begin{minipage}[t][5cm][t]{0.4\linewidth} 
    \centering
    \includegraphics[width=3cm, height=3cm]{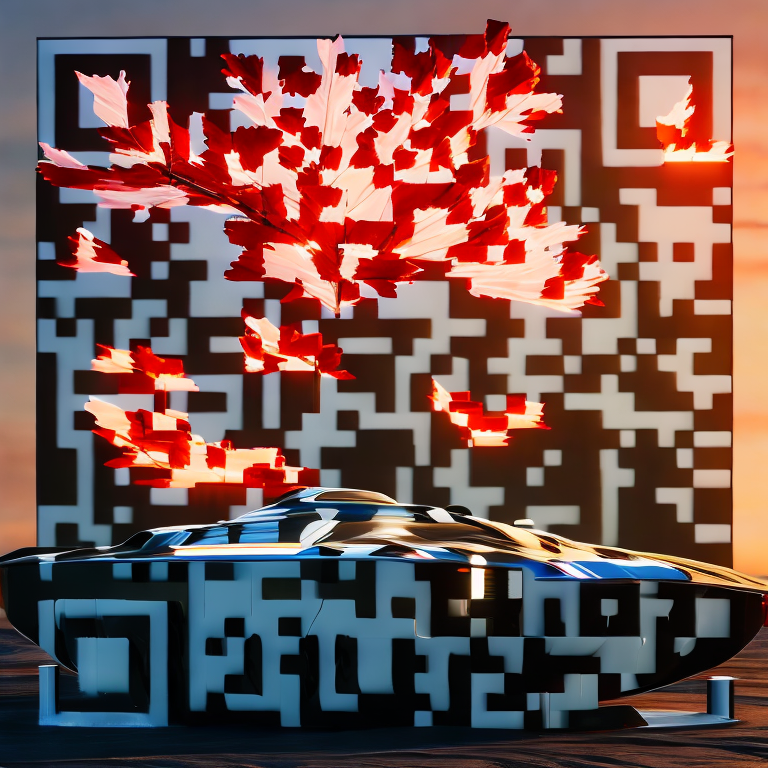}
    \subcaption{\small Generated QR code}
    \label{fig:4b}
  \end{minipage}
  \vspace{-5em}
  \caption{Comparison of input and output for Music2P's QR code generation service.~[Best viewed in color]}
  \label{fig4}
  \vspace{-1.5em}
\end{figure}

The QR code generation feature in Music2P allows users to create visually integrated QR codes within their album covers. The process involves the following steps:

\begin{enumerate}
    \item The user provides a base image and a QR code.
    \item The user inputs text describing the desired aesthetic.
    \item The system generates an AI-enhanced QR code integrated with the base image.
\end{enumerate}
For example, if a user provides the base image shown in \cref{fig:4a} and specifies ``realistic, 8K'' as the text input along with techno music, Music2P's QR code service will produce an output as illustrated in \cref{fig:4b}.
It is important to note that the aesthetic quality and functionality of the output QR code may be compromised if the user submits a personal photograph or an image containing patterned objects. 
The QR code generation model utilizes several key hyperparameters that significantly influence the output:

\begin{itemize}
    \item Guidance scale: Higher values result in sharper final images, affecting both the QR code and the base image.
    \item ControlNet conditioning scale: This parameter adjusts the balance between the QR code and the base image. Values range from 0 to 5, where lower values result in a less visible QR code, and higher values make the QR code more prominent.
    \item Strength: This parameter also controls the visibility of the QR code over the base image. Lower values result in a subtle QR code, while higher values make it more dominant.
\end{itemize}
These hyperparameters allow for fine-tuning the visual integration and readability of the QR code within the album cover design.

After we trained the model to generate QR codes and album covers, we utilized Ngrok to avoid running the code to use our service. Ngrok provides the IP address and the port number to use our service. After the user uploads the image, mp3 file, and the style for the album cover on the website generated by Ngrok, the user can get the album cover without running our code.

%% file: section/3_demo.tex
\section{System Implementation \& User Guide}
\label{sec:implementation_and_guide}
Music2P is an open-source web system solution accessible to anyone\footnote{\url{https://github.com/JC-78/Music2P}}. The system is designed to be user-friendly, requiring minimal technical knowledge for deployment and operation. To utilize Music2P, users need to follow these steps:

\begin{enumerate}
    \item Execute all cells in the provided Colab notebook.
    \item Copy the URL generated in the final cell into the \texttt{url} variable in \texttt{app.py}.
    \item Run \texttt{app.py} and access the generated endpoint/link of the solution.
    \item Through the user interface, upload the following:
    \begin{itemize}
        \item An MP3 file of the music for which the album cover will be generated.
        \item An image to serve as a foundation or reference for the album cover art.
        \item A text description specifying the desired tone or style for the album cover.
    \end{itemize}
    \item Receive the AI-generated album cover within approximately one to two minutes.
\end{enumerate}
Our implementation utilizes Flask~\cite{grinberg2018flask} to connect to a free-tier Google Colab notebook with GPU runtime. By using ngrok, as long as an individual musician or one member of a team uses google Colab, we can eliminate any need for users to be on the same network as the server or any low-level technical knowledge. Therefore, with Music2P, we hope to help musicians and small companies with their marketing and promotion strategies by minimizing the cost and time needed to leverage AI for album cover generation.

%% file: section/4_conclusion.tex
\section{Conclusion \& Future Work}
This paper introduces Music2P, a multi-modal system for automated album cover generation that integrates advanced AI techniques including BLIP for image captioning, LP-music-caps for music-to-text conversion, LoRA for image segmentation, and ControlNet for album cover and QR code generation.
We've also explored a cost-effective deployment using Ngrok, making the solution accessible to a wider range of users.
For future directions, we would need to focus on scalable infrastructure deployment.
Additionally, we need to train more customized LoRA for our Music2P solution.
This is essential because when generating album covers, if user input images consist of faces or objects of pattern, the aesthetic quality of output deteriorates significantly.

\section*{Acknowledgments}
This work was supported by the Institute of Information \& Communications Technology Planning \& Evaluation (IITP) grant funded by the Korea government (MSIT) (RS-2022-00143911, AI Excellence Global Innovative Leader Education Program).